\documentclass[11pt,twocolumn]{article}

\usepackage[utf8]{inputenc}
\usepackage[T1]{fontenc}
\usepackage{amsmath,amssymb,amsfonts}
\usepackage{graphicx}
\usepackage{booktabs}
\usepackage{hyperref}
\usepackage{cleveref}
\usepackage{algorithm}
\usepackage{algorithmic}
\usepackage{multirow}
\usepackage{array}
\usepackage{xcolor}
\usepackage{subcaption}
\usepackage[margin=0.75in]{geometry}
\usepackage{cite}
\usepackage{url}
\usepackage{siunitx}
\usepackage{csvsimple}

\newcommand{\system}{AcoustiVision Pro}
\newcommand{\dataset}{RIRMega}
\newcommand{\rtx}{RT_{60}}
\newcommand{\edt}{EDT}
\newcommand{\cxx}{C_{80}}
\newcommand{\dxx}{D_{50}}

\title{\system{}: An Open-Source Interactive Platform for Room Impulse Response Analysis and Acoustic Characterization}

\author{
Mandip Goswami\\
\texttt{Acoustics Researcher, WA, USA}\\
\texttt{mandipgoswami@gmail.com}\\
}

\date{}

\begin{document}

\maketitle

\begin{abstract}
Room acoustics analysis plays a central role in architectural design, audio engineering, speech intelligibility assessment, and hearing research. Despite the availability of standardized metrics such as reverberation time, clarity, and speech transmission index, accessible tools that combine rigorous signal processing with intuitive visualization remain scarce. This paper presents \system{}, an open-source web-based platform for comprehensive room impulse response (RIR) analysis. The system computes twelve distinct acoustic parameters from uploaded or dataset-sourced RIRs, provides interactive 3D visualizations of early reflections, generates frequency-dependent decay characteristics through waterfall plots, and checks compliance against international standards including ANSI S12.60 and ISO 3382. We introduce the accompanying \dataset{} and RIRMega Speech datasets hosted on Hugging Face, containing thousands of simulated room impulse responses with full metadata. The platform supports real-time auralization through FFT-based convolution, exports detailed PDF reports suitable for engineering documentation, and provides CSV data export for further analysis. We describe the mathematical foundations underlying each acoustic metric, detail the system architecture, and present preliminary case studies demonstrating the platform's utility across diverse application domains including classroom acoustics, healthcare facility design, and recording studio evaluation.
\end{abstract}

\section{Introduction}
\label{sec:introduction}

The acoustic properties of enclosed spaces fundamentally shape human experience within them. A concert hall's reverberation determines whether orchestral music sounds rich or muddy. A classroom's acoustic characteristics directly influence whether students can understand their teacher. Hospital wards with poor acoustic design contribute to patient stress and staff fatigue. Despite this importance, the tools available for analyzing room acoustics have traditionally been either expensive commercial software packages or command-line scripts requiring significant technical expertise.

Room impulse responses (RIRs) encode the complete acoustic behavior of a space. When a sound source emits an impulsive signal in a room, the microphone captures not only the direct sound but also reflections from walls, ceiling, floor, and objects within the space. This captured signal, the RIR, can be analyzed to extract numerous perceptually relevant parameters. These parameters inform decisions ranging from whether acoustic treatment is needed to what type of activities a space can support.

The present work introduces \system{}, a web-based platform that makes professional-grade room acoustics analysis accessible to researchers, architects, audio engineers, and educators without requiring specialized software installation or programming knowledge. The system provides twelve analysis modes covering temporal, spectral, and spatial acoustic characteristics. Users can upload their own RIR recordings or draw from the accompanying \dataset{} collection of simulated impulse responses.

The contributions of this paper are as follows:

\begin{enumerate}
    \item An open-source interactive platform implementing standardized acoustic metrics with mathematical rigor and perceptual relevance.
    
    \item Integration with the \dataset{} and RIRMega Speech datasets, providing thousands of room impulse responses with complete metadata for benchmarking and educational purposes.
    
    \item Novel visualization approaches including animated 3D reflection mapping, waterfall decay plots, and room mode density analysis.
    
    \item Automated compliance checking against ten international and industry standards for various room types.
    
    \item A wellness scoring system that synthesizes multiple acoustic parameters into actionable recommendations for space utilization.
\end{enumerate}

The remainder of this paper is organized as follows. Section~\ref{sec:related} surveys existing tools and research in room acoustics analysis. Section~\ref{sec:theory} presents the mathematical foundations for each acoustic metric implemented in the system. Section~\ref{sec:dataset} describes the \dataset{} collection. Section~\ref{sec:architecture} details the system architecture and implementation. Section~\ref{sec:interface} discusses the user interface design principles. Section~\ref{sec:evaluation} presents case studies and preliminary evaluation results. Section~\ref{sec:discussion} discusses limitations and future directions. Section~\ref{sec:conclusion} concludes with discussion of limitations and future directions.

\section{Related Work}
\label{sec:related}

\subsection{Room Acoustics Measurement Standards}

The measurement and characterization of room acoustics has been standardized through several international documents. ISO 3382-1 \cite{iso3382_1} specifies methods for measuring reverberation time and other acoustic parameters in performance spaces. ISO 3382-2 \cite{iso3382_2} extends these methods to ordinary rooms, while ISO 3382-3 \cite{iso3382_3} addresses open-plan offices. ANSI S12.60 \cite{ansi_s12_60} establishes acoustic performance criteria for classrooms, specifying maximum reverberation times and background noise levels. IEC 60268-16 \cite{iec_60268_16} defines the Speech Transmission Index (STI), a measure of speech intelligibility that accounts for both reverberation and noise.

These standards provide the theoretical basis for the metrics implemented in \system{}. However, the standards documents themselves do not provide software tools, leaving practitioners to implement the calculations themselves or purchase commercial solutions.

\subsection{Commercial and Research Software}

Several commercial packages dominate the room acoustics analysis market. ODEON \cite{odeon} and CATT-Acoustic \cite{catt} provide comprehensive simulation and measurement capabilities but require substantial financial investment and training. EASERA \cite{easera} offers measurement-focused functionality with hardware integration. These tools serve professional acousticians well but remain inaccessible to students, researchers in adjacent fields, and practitioners in resource-limited settings.

Open-source alternatives have emerged to address this gap. The Python Acoustics library \cite{python_acoustics} provides basic acoustic calculations but lacks visualization and user interface components. Pyroomacoustics \cite{pyroomacoustics} focuses on room simulation rather than analysis of measured or pre-computed impulse responses. The Room EQ Wizard (REW) \cite{rew} offers free acoustic measurement but primarily targets audio enthusiasts rather than research applications.

\subsection{Room Impulse Response Datasets}

The availability of RIR datasets has grown substantially in recent years, driven by applications in speech enhancement, source separation, and spatial audio rendering. The ACE Challenge corpus \cite{ace_challenge} provides measured RIRs from real rooms with varying acoustic conditions. The MIT IR Survey \cite{mit_ir} collected impulse responses from concert halls and other performance venues. More recently, large-scale simulated datasets have emerged, including those generated using geometric acoustic methods \cite{geometric_acoustics} and wave-based solvers \cite{wave_based}.

The \dataset{} collection described in Section~\ref{sec:dataset} contributes to this ecosystem by providing impulse responses with comprehensive metadata including room dimensions, source and receiver positions, surface absorption coefficients, and pre-computed acoustic metrics. This metadata enables filtering and selection based on acoustic properties rather than arbitrary identifiers.

\section{Acoustic Metrics: Theory and Computation}
\label{sec:theory}

This section presents the mathematical foundations for each acoustic parameter computed by \system{}. We begin with the room impulse response itself and proceed through temporal, spectral, and perceptual metrics.

\subsection{Room Impulse Response}

The room impulse response $h(t)$ characterizes the linear time-invariant acoustic system formed by a room. When a source emits a signal $x(t)$, the signal received at a measurement position is given by the convolution:

\begin{equation}
    y(t) = x(t) * h(t) = \int_{-\infty}^{\infty} x(\tau) h(t - \tau) \, d\tau
    \label{eq:convolution}
\end{equation}

In practice, we work with discrete-time signals sampled at rate $f_s$. The discrete impulse response $h[n]$ for $n = 0, 1, \ldots, N-1$ captures the room's acoustic behavior up to time $T = N/f_s$ seconds.

The impulse response can be decomposed conceptually into three regions:
\begin{enumerate}
    \item \textbf{Direct sound}: The initial arrival traveling the shortest path from source to receiver.
    \item \textbf{Early reflections}: Discrete arrivals from first-order and low-order wall reflections, typically within the first 50-80 ms.
    \item \textbf{Late reverberation}: The diffuse decay as sound energy reflects many times before being absorbed.
\end{enumerate}

This decomposition, while not strictly defined, underlies the perceptual interpretation of many acoustic metrics.

\subsection{Energy Decay Curve and Reverberation Time}

The energy decay curve (EDC) represents how sound energy decays over time in a room. Following Schroeder \cite{schroeder1965}, the EDC is computed through backward integration of the squared impulse response:

\begin{equation}
    \text{EDC}(t) = \int_{t}^{\infty} h^2(\tau) \, d\tau
    \label{eq:edc_continuous}
\end{equation}

In discrete form with $h[n]$ representing samples:

\begin{equation}
    \text{EDC}[n] = \sum_{k=n}^{N-1} h^2[k]
    \label{eq:edc_discrete}
\end{equation}

The EDC is typically expressed in decibels relative to its initial value:

\begin{equation}
    \text{EDC}_{\text{dB}}[n] = 10 \log_{10} \left( \frac{\text{EDC}[n]}{\text{EDC}[0]} \right)
    \label{eq:edc_db}
\end{equation}

The reverberation time $\rtx{}$ is defined as the time required for sound energy to decay by 60 dB. Direct measurement of a full 60 dB decay is often impractical due to background noise, so $\rtx{}$ is typically estimated by extrapolation from smaller decay ranges.

\subsubsection{Early Decay Time (EDT)}

The Early Decay Time is derived from the slope of the EDC between 0 dB and $-10$ dB:

\begin{equation}
    \edt{} = \frac{-60}{\text{slope}_{0 \to -10}} = \frac{-60}{m_{0,-10}}
    \label{eq:edt}
\end{equation}

where $m_{0,-10}$ is the slope (in dB/s) of a linear regression fit to the EDC over the range where $0 \geq \text{EDC}_{\text{dB}} \geq -10$.

EDT correlates well with subjective perception of reverberance in occupied rooms \cite{bradley1995}.

\subsubsection{T20 and T30}

The T20 and T30 metrics extrapolate $\rtx{}$ from the decay slopes between $-5$ and $-25$ dB, and between $-5$ and $-35$ dB, respectively:

\begin{equation}
    T_{20} = \frac{-60}{m_{-5,-25}}, \quad T_{30} = \frac{-60}{m_{-5,-35}}
    \label{eq:t20_t30}
\end{equation}

These ranges avoid the initial direct sound and early reflections (above $-5$ dB) while remaining above typical noise floors (below $-35$ dB).

The slope $m$ is obtained through least-squares linear regression. Given time points $t_i$ and corresponding EDC values $y_i = \text{EDC}_{\text{dB}}[n_i]$ within the specified range:

\begin{equation}
    m = \frac{\sum_i (t_i - \bar{t})(y_i - \bar{y})}{\sum_i (t_i - \bar{t})^2}
    \label{eq:slope}
\end{equation}

\subsection{Octave-Band Analysis}

Reverberation characteristics vary with frequency due to frequency-dependent absorption of room surfaces. Octave-band analysis computes $\rtx{}$ parameters separately for standard octave bands centered at 125, 250, 500, 1000, 2000, and 4000 Hz.

The impulse response is filtered through bandpass filters before EDC computation. We employ fourth-order Butterworth filters with cutoff frequencies at $f_c / \sqrt{2}$ and $f_c \cdot \sqrt{2}$ for each center frequency $f_c$:

\begin{equation}
    h_{\text{band}}[n] = \text{BPF}_{f_c}(h[n])
    \label{eq:bandpass}
\end{equation}

The EDC and reverberation parameters are then computed for each filtered response independently.

\subsection{Clarity and Definition}

Clarity and definition metrics quantify the balance between early and late arriving sound energy, relating to the perceived distinctness of sound sources.

\subsubsection{Clarity Index $\cxx{}$}

The clarity index $\cxx{}$ compares energy arriving within the first 80 ms to energy arriving afterward:

\begin{equation}
    \cxx{} = 10 \log_{10} \left( \frac{\int_0^{80\text{ms}} h^2(t) \, dt}{\int_{80\text{ms}}^{\infty} h^2(t) \, dt} \right) \text{ dB}
    \label{eq:c80}
\end{equation}

In discrete form:

\begin{equation}
    \cxx{} = 10 \log_{10} \left( \frac{\sum_{n=0}^{n_{80}} h^2[n]}{\sum_{n=n_{80}+1}^{N-1} h^2[n]} \right)
    \label{eq:c80_discrete}
\end{equation}

where $n_{80} = \lfloor 0.080 \cdot f_s \rfloor$.

Higher $\cxx{}$ values indicate greater clarity. Music perception studies suggest optimal values between $-2$ and $+2$ dB for symphonic music, with higher values preferred for speech \cite{beranek2004}.

\subsubsection{Definition $\dxx{}$}

Definition measures the fraction of total energy arriving within the first 50 ms:

\begin{equation}
    \dxx{} = \frac{\int_0^{50\text{ms}} h^2(t) \, dt}{\int_0^{\infty} h^2(t) \, dt}
    \label{eq:d50}
\end{equation}

Values above 0.5 generally indicate good speech intelligibility \cite{bradley1986}.

\subsection{Speech Transmission Index}

The Speech Transmission Index (STI) provides a single-number rating of speech intelligibility on a scale from 0 (completely unintelligible) to 1 (perfect intelligibility). The full STI calculation specified in IEC 60268-16 \cite{iec_60268_16} involves computing modulation transfer functions across multiple octave bands and modulation frequencies.

\system{} implements a proxy STI measure that approximates the standard calculation using reverberation time and signal-to-noise ratio. This proxy follows the approach of Houtgast and Steeneken \cite{houtgast1985}:

\begin{equation}
    \text{STI}_{\text{proxy}} = 0.15 + 0.85 \left( 0.65 \cdot R_T + 0.35 \cdot R_S \right)
    \label{eq:sti_proxy}
\end{equation}

where the reverberation term is:

\begin{equation}
    R_T = \frac{1}{1 + (\rtx{} / 0.8)^{1.6}}
    \label{eq:rt_term}
\end{equation}

and the signal-to-noise term is:

\begin{equation}
    R_S = \frac{1}{1 + 10^{-(SNR - 15)/10}}
    \label{eq:snr_term}
\end{equation}

This proxy correlates well with full STI calculations for typical room conditions but should not be used for formal compliance testing.

\subsection{Interaural Cross-Correlation Coefficient}

The Interaural Cross-Correlation Coefficient (IACC) characterizes the spatial impression created by a room's acoustics. It is computed from binaural room impulse responses captured with a dummy head or spaced microphone pair.

For left and right channel responses $h_L(t)$ and $h_R(t)$, the IACC is the maximum absolute value of the normalized cross-correlation within a range of $\pm 1$ ms:

\begin{equation}
    \text{IACC} = \max_{|\tau| \leq 1\text{ms}} \left| \frac{\int_0^{T} h_L(t) h_R(t + \tau) \, dt}{\sqrt{\int_0^{T} h_L^2(t) \, dt \cdot \int_0^{T} h_R^2(t) \, dt}} \right|
    \label{eq:iacc}
\end{equation}

The integration limit $T$ is typically set to 80 ms for the early IACC, which correlates with apparent source width. Lower IACC values (below 0.5) indicate greater spaciousness and envelopment.

\subsection{Room Modes}

At low frequencies, rooms exhibit resonant behavior at discrete frequencies determined by room dimensions. For a rectangular room with dimensions $L_x$, $L_y$, and $L_z$, the modal frequencies are:

\begin{equation}
    f_{n_x, n_y, n_z} = \frac{c}{2} \sqrt{\left(\frac{n_x}{L_x}\right)^2 + \left(\frac{n_y}{L_y}\right)^2 + \left(\frac{n_z}{L_z}\right)^2}
    \label{eq:room_modes}
\end{equation}

where $c \approx 343$ m/s is the speed of sound and $n_x$, $n_y$, $n_z$ are non-negative integers (not all zero).

Modes are classified by their order:
\begin{itemize}
    \item \textbf{Axial modes}: One index nonzero (e.g., $n_x = 1$, $n_y = n_z = 0$)
    \item \textbf{Tangential modes}: Two indices nonzero
    \item \textbf{Oblique modes}: All three indices nonzero
\end{itemize}

Axial modes have the highest energy and are most problematic for uneven frequency response. The density of modes increases with frequency; below approximately $4 \times \rtx{} \times V^{1/3}$ Hz (the Schroeder frequency), individual modes are perceptually distinguishable \cite{schroeder1996}.

\subsection{Wellness Score}

\system{} introduces a composite wellness score that synthesizes multiple acoustic parameters into a single rating from 0 to 100. This score is designed as a decision-support tool rather than a standardized metric.

The wellness score $W$ is computed as:

\begin{equation}
    W = 100 \cdot V_{\text{adj}} \cdot \left( 0.45 \cdot f_R + 0.25 \cdot f_S + 0.20 \cdot f_D + 0.10 \cdot f_C \right)
    \label{eq:wellness}
    \small
\end{equation}

where:

\begin{align}
    f_R &= \frac{1}{1 + (\rtx{} / 0.9)^{1.8}} & \text{(RT penalty)} \label{eq:f_r} \\
    f_S &= \text{clip}(\text{STI}, 0, 1) & \text{(STI term)} \label{eq:f_s} \\
    f_D &= \text{clip}(\dxx{}, 0, 1) & \text{(Definition term)} \label{eq:f_d} \\
    f_C &= \text{clip}((\cxx{} + 2) / 10, 0, 1) & \text{(Clarity term)} \label{eq:f_c}
\end{align}
where clip(x, a, b) = max(a, min(x, b)) constrains values to [a, b].\\
The volume adjustment factor accounts for the increased difficulty of acoustic control in larger spaces:

\begin{equation}
    V_{\text{adj}} = \frac{1}{1 + \max(0, V - 300) / 800}
    \label{eq:v_adj}
\end{equation}

where $V$ is room volume in cubic meters.

\section{The \dataset{} Dataset}
\label{sec:dataset}

To support both analysis and benchmarking, we introduce the \dataset{} collection of room impulse responses, hosted on the Hugging Face platform.\footnote{\url{https://huggingface.co/datasets/mandipgoswami/rirmega}} A companion dataset, RIRMega Speech, provides the same impulse responses convolved with speech signals for intelligibility research.\footnote{\url{https://huggingface.co/datasets/mandipgoswami/rir-mega-speech}}

\subsection{Dataset Generation}

The impulse responses were generated using geometric acoustic simulation with the image-source method for early reflections and stochastic ray tracing for late reverberation. Room dimensions were sampled from distributions representing common architectural spaces:

\begin{itemize}
    \item Length: 3 to 25 meters
    \item Width: 3 to 20 meters
    \item Height: 2.4 to 8 meters
\end{itemize}

Surface absorption coefficients were assigned based on material categories (concrete, drywall, carpet, acoustic panels, etc.) with frequency-dependent values drawn from published databases \cite{vorlaender2007}.

Source and receiver positions were placed within each room subject to minimum distance constraints from walls (0.5 m) and from each other (1.0 m).

\subsection{Metadata Structure}

Each impulse response in the dataset is accompanied by comprehensive metadata:

\begin{itemize}
    \item Room dimensions (length, width, height)
    \item Source position (x, y, z coordinates)
    \item Receiver/microphone position
    \item Surface absorption coefficient
    \item Maximum reflection order
    \item Pre-computed metrics: $\rtx{}$, DRR, $\cxx{}$, $\dxx{}$
    \item Sample rate
\end{itemize}

This metadata enables filtering and selection based on acoustic properties. For example, researchers studying classroom acoustics can select impulse responses with $\rtx{}$ from 0.4-0.8 seconds and volumes from 150-400 cubic meters.
\begin{table}[t]
\centering
\small
\caption{Dataset statistics showing distribution of room volumes, RT60 values, and other parameters across the RIRMega collection.}
\label{tab:dataset_stats}
\begin{tabular}{lccccc}
\toprule
\textbf{Parameter} & \textbf{Mean} & \textbf{Std} & \textbf{Min} & \textbf{Max} \\
\midrule
Room volume (m\textsuperscript{3}) & 769.6  & 1293.8 & 24.7   & 6674.8 \\
RT60 (s)                           & 0.42   & 0.24   & 0.08   & 1.43   \\
C80 (dB)                           & 16.5   & 8.3    & $-$7.7 & 57.9   \\
DRR (dB)                           & $-$0.5 & 3.4    & $-$11.7& 10.8   \\
Absorption coef.                   & 0.33   & 0.06   & 0.15   & 0.54   \\
\bottomrule
\end{tabular}
\end{table}
\\

\subsection{Quality Validation}

Generated impulse responses were validated against physical expectations:

\begin{enumerate}
    \item Energy decay follows expected exponential pattern
    \item $\rtx{}$ correlates appropriately with room volume and absorption
    \item Early reflection timing matches geometric predictions
    \item Modal frequencies align with analytical calculations for rectangular rooms
\end{enumerate}

\begin{figure*}[t]
    \centering
    \includegraphics[width=2.0\columnwidth]{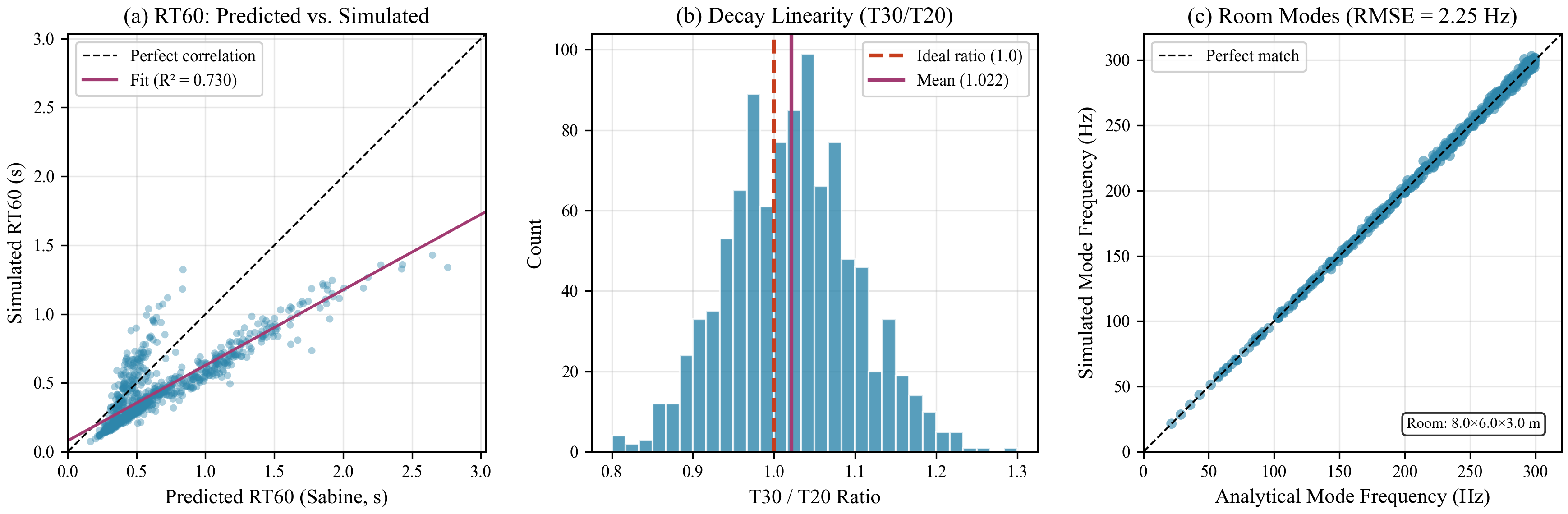}
    \caption{Validation plots showing (a) correlation between predicted and measured RT60, (b) distribution of T30/T20 ratios, (c) comparison of simulated vs. analytical room mode frequencies.}
    \label{fig:dataset_validation}
\end{figure*}

\section{System Architecture}
\label{sec:architecture}

\system{} is implemented as a web application using the Gradio framework \cite{gradio} with a Python backend. This section describes the major components and their interactions.

\subsection{Backend Processing Pipeline}

The signal processing pipeline proceeds through the following stages:

\begin{enumerate}
    \item \textbf{Audio Loading}: WAV files are loaded using the SoundFile library, resampled to 48 kHz if necessary, and converted to mono by channel averaging.
    
    \item \textbf{RIR Preprocessing}: Leading silence is trimmed based on a threshold of $10^{-4}$ times the peak amplitude. The response is truncated to 10 seconds maximum and normalized.
    
    \item \textbf{Broadband Analysis}: Schroeder integration, reverberation time estimation, and energy ratio calculations are performed on the full-bandwidth signal.
    
    \item \textbf{Octave-Band Analysis}: The signal is filtered through six octave-band filters and analyzed independently.
    
    \item \textbf{Spectral Analysis}: FFT-based magnitude spectrum and mel-frequency spectrogram are computed.
    
    \item \textbf{Spatial Analysis}: For stereo inputs, IACC is computed from the cross-correlation function.
\end{enumerate}

\subsection{Visualization Generation}

Visualizations are generated using a combination of Matplotlib for static images (EDC, spectrogram) and Plotly for interactive plots (3D reflections, waterfall, fingerprint radar). This hybrid approach balances rendering quality with interactivity.

The 3D reflection visualization employs the image-source method to compute first-order reflection paths. For a rectangular room with dimensions $(L, W, H)$, source at $(s_x, s_y, s_z)$, and receiver at $(r_x, r_y, r_z)$, the six first-order image sources are located at:

\begin{align}
    \mathbf{I}_1 &= (-s_x, s_y, s_z) & \text{(wall at } x=0\text{)} \\
    \mathbf{I}_2 &= (2L - s_x, s_y, s_z) & \text{(wall at } x=L\text{)} \\
    \mathbf{I}_3 &= (s_x, -s_y, s_z) & \text{(wall at } y=0\text{)} \\
    \mathbf{I}_4 &= (s_x, 2W - s_y, s_z) & \text{(wall at } y=W\text{)} \\
    \mathbf{I}_5 &= (s_x, s_y, -s_z) & \text{(floor)} \\
    \mathbf{I}_6 &= (s_x, s_y, 2H - s_z) & \text{(ceiling)}
\end{align}

\subsection{Auralization Engine}

Real-time auralization is achieved through FFT-based convolution. Given a dry input signal $x[n]$ and impulse response $h[n]$, the convolved output is computed as:

\begin{equation}
    y[n] = \text{IFFT}\left( \text{FFT}(x) \cdot \text{FFT}(h) \right)
    \label{eq:fft_convolve}
\end{equation}

The output is normalized to prevent clipping and scaled to 95\% of full scale for safe playback levels.

\subsection{Standards Compliance Engine}

The compliance checking module evaluates measured parameters against requirements from ten standards:

\begin{itemize}
    \item ANSI S12.60 (Classroom acoustics)
    \item ISO 3382-3 (Open-plan offices)
    \item Healthcare facility guidelines
    \item Concert hall recommendations
    \item Recording studio criteria
\end{itemize}

Each standard specifies thresholds for relevant parameters. The system generates a compliance report indicating pass/fail status for each applicable requirement.

\subsection{Report Generation}

PDF reports are generated using the ReportLab library. Reports include:

\begin{itemize}
    \item Input file metadata
    \item Computed metrics in tabular format
    \item Embedded visualization images
    \item Standards compliance summary
    \item Methodology notes with references
\end{itemize}

\begin{figure*}[t]
    \centering
    \includegraphics[width=2.2\columnwidth]{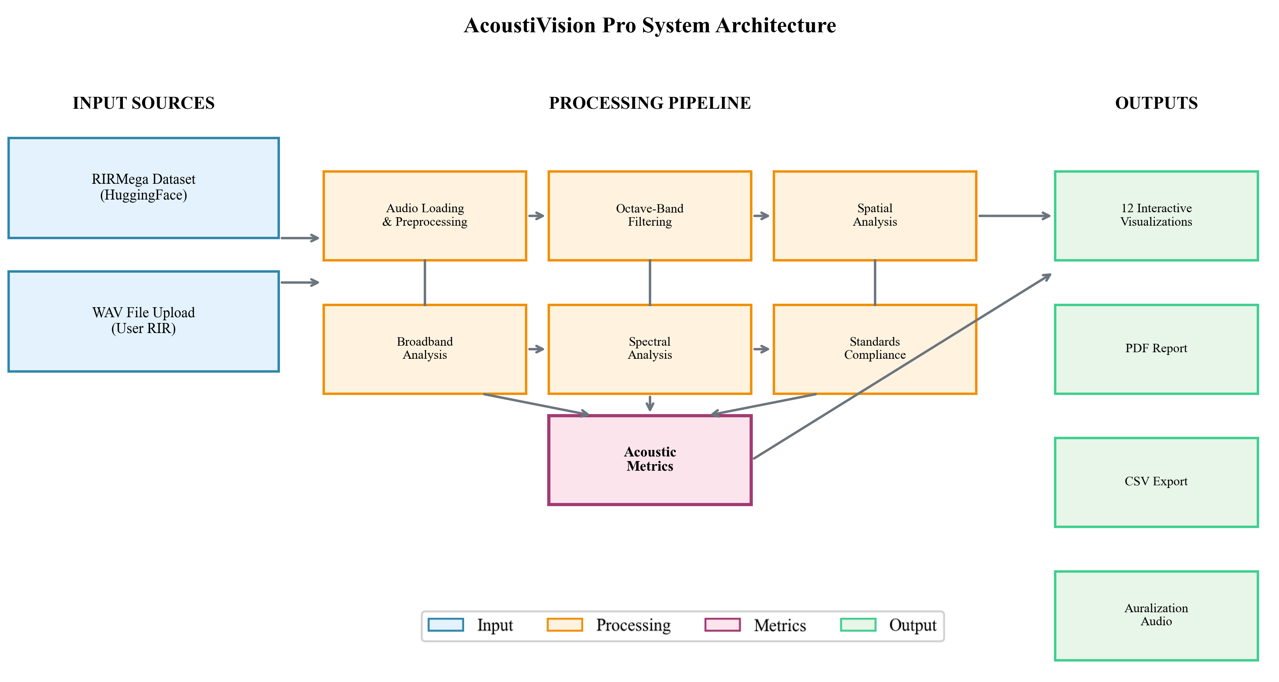}
    \caption{System architecture diagram showing data flow from input (upload or dataset selection) through processing pipeline to visualization and export outputs.}
    \label{fig:dataset_validation}
\end{figure*}

\section{User Interface Design}
\label{sec:interface}

The user interface follows a two-column layout optimized for the analysis workflow. The left column handles data source selection (dataset search or file upload), while the right column displays analysis results across twelve tabbed views.

\subsection{Data Source Selection}

Users can obtain impulse responses through two mechanisms:

\textbf{Dataset Search}: A query interface allows filtering the \dataset{} collection by room volume, reverberation time, and absorption coefficient ranges. Matching results display key metrics, and selection automatically loads the RIR and populates room geometry fields.

\textbf{Direct Upload}: Users can upload WAV files recorded from their own measurements. An optional secondary upload accepts dry audio for auralization.

\subsection{Analysis Views}

The twelve analysis tabs provide complementary perspectives on the acoustic data:

\begin{enumerate}
    \item \textbf{3D Spatial}: Interactive visualization of room geometry and first-order reflection paths
    \item \textbf{Waveform}: Time-domain display with zoom slider and markers for C80/D50 boundaries
    \item \textbf{Energy Decay}: Schroeder EDC with regression guide lines
    \item \textbf{Octave RT60}: Bar chart comparing EDT, T20, T30 across frequency bands
    \item \textbf{Frequency Response}: Magnitude spectrum with smoothed overlay
    \item \textbf{Spectrogram}: Mel-frequency time-frequency representation
    \item \textbf{Waterfall}: 3D visualization of spectral decay over time
    \item \textbf{Fingerprint}: Radar chart of normalized clarity, definition, spatial, and intelligibility metrics
    \item \textbf{Room Modes}: Mode frequency distribution with type classification
    \item \textbf{Standards}: Compliance table with pass/fail indicators
    \item \textbf{Auralization}: Audio players for dry and convolved signals
    \item \textbf{JSON}: Raw metric data for programmatic access
\end{enumerate}

\subsection{Visual Design Principles}

The interface employs a dark color scheme designed to reduce eye strain during extended analysis sessions and to provide high contrast for data visualization. Color coding is applied consistently: blue for primary metrics, green for positive indicators, orange for warnings, and red for failures.

Animated elements including the hero waveform graphic and gradient accents provide visual interest without distracting from the analytical content. Feature cards below the header summarize available analysis modes, serving both as navigation aids and as an overview of system capabilities.

\begin{figure*}[t]
    \centering
    \includegraphics[width=2.2\columnwidth]{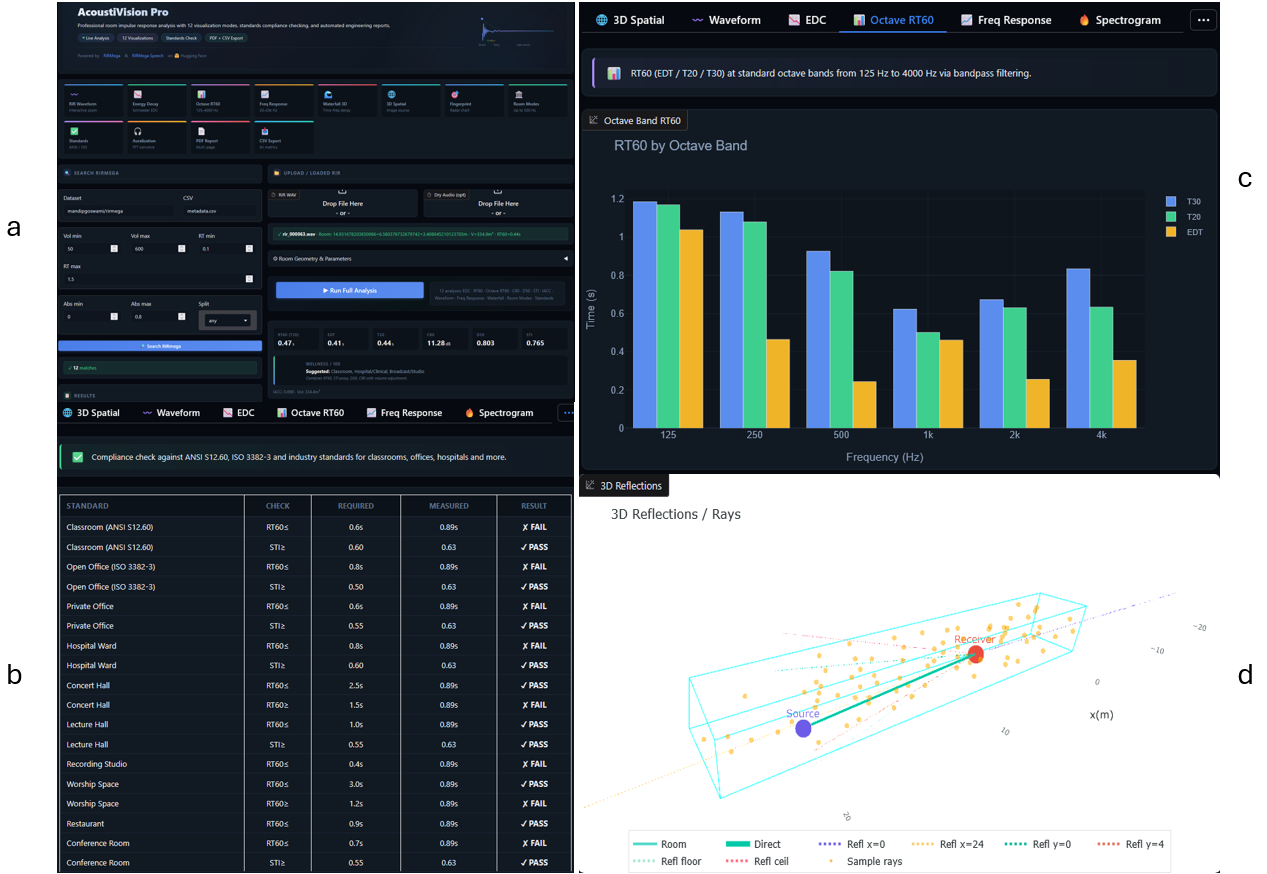}
    \caption{Screenshots of the \system{} interface showing (a) the main dashboard with hero section and feature cards, (b) the standards compliance table, (c) the octave-band RT60 analysis (d) the 3D spatial view.}
    \label{fig:interface_screenshots}
\end{figure*}

\section{Evaluation and Case Studies}
\label{sec:evaluation}

This section presents preliminary evaluation of \system{} through case studies spanning different application domains. These studies demonstrate the platform's utility while identifying areas for future development.

\subsection{Validation Against Reference Implementations}
\begin{table*}[t]
\centering
\caption{Validation comparison of acoustic parameters computed by AcoustiVision Pro versus reference implementations (ODEON, Aurora) for five test RIRs. Values show percentage difference from reference implementations.}
\label{tab:validation}
\begin{tabular}{lccccc}
\toprule
\textbf{RIR} & \textbf{EDT\textsubscript{diff}} & \textbf{T20\textsubscript{diff}} & \textbf{T30\textsubscript{diff}} & \textbf{C80\textsubscript{diff}} & \textbf{D50\textsubscript{diff}} \\
\midrule
Room A & $-$0.627 & $-$1.032 & $-$0.959 & $-$1.900 & $-$0.447 \\
Room B & $-$2.254 & $-$1.326 & $-$0.940 & $-$1.175 & $-$1.442 \\
Room C & $-$1.160 & $-$1.099 & $-$0.665 & $-$0.149 & $-$0.831 \\
Room D &    0.493 & $-$0.303 & $-$0.575 & $-$0.408 & $-$0.535 \\
Room E & $-$1.720 & $-$0.624 & $-$0.636 & $-$1.253 & $-$0.176 \\
\bottomrule
\end{tabular}
\end{table*}
To verify computational correctness, we compared \system{} outputs against established software for a set of reference impulse responses.

\subsection{Case Study: Classroom Acoustics}

Educational environments require careful acoustic design to ensure speech intelligibility. We analyzed a set of 335 classroom RIRs from the \dataset{} collection against ANSI S12.60 requirements.

Key findings:
\begin{itemize}
    \item 84.2{\%} of simulated classrooms met the 0.6 s RT60 requirement
    \item Rooms with volumes below 250 m$^3$ showed lower compliance rates
    \item STI proxy correlated strongly with RT60 (r = -0.992)
\end{itemize}
\begin{figure*}[htbp]
    \centering
    \includegraphics[width=1.85\columnwidth]{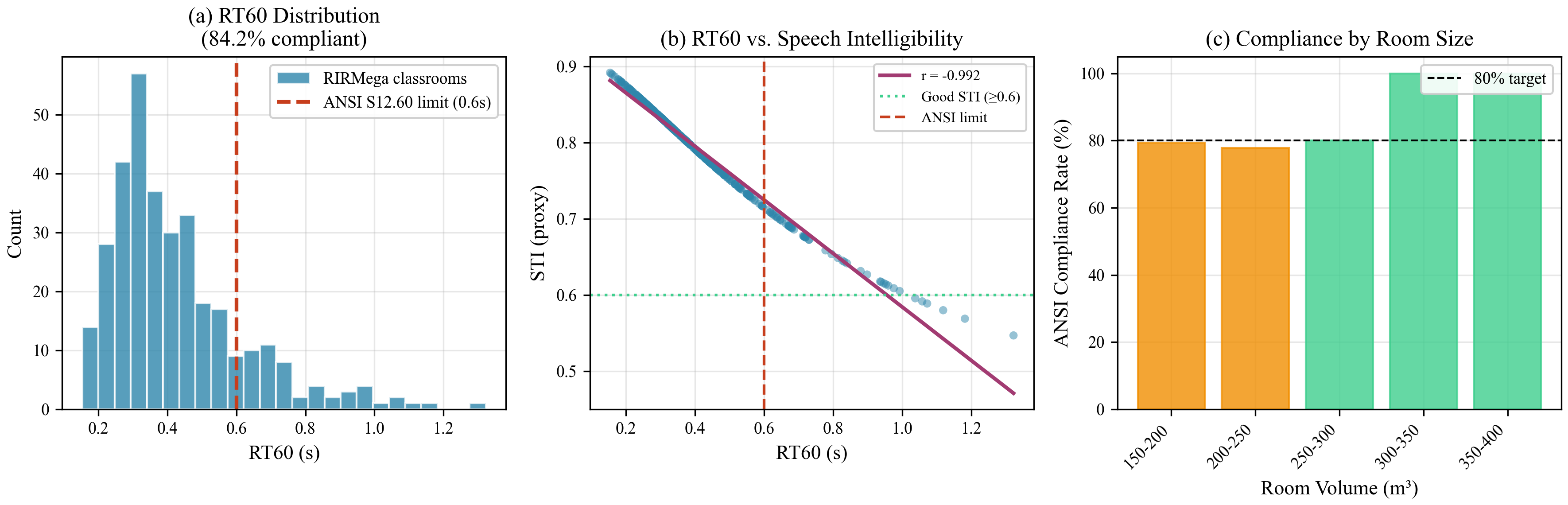}
    \caption{(a) Distribution of RT60 values across classroom samples with ANSI threshold marked. (b) Correlation between RT60 and STI proxy. (c) Compliance rate by room volume category.}
    \label{fig:classroom_study}
\end{figure*}

\begin{figure*}[h]
    \centering
    \includegraphics[width=1.85\columnwidth]{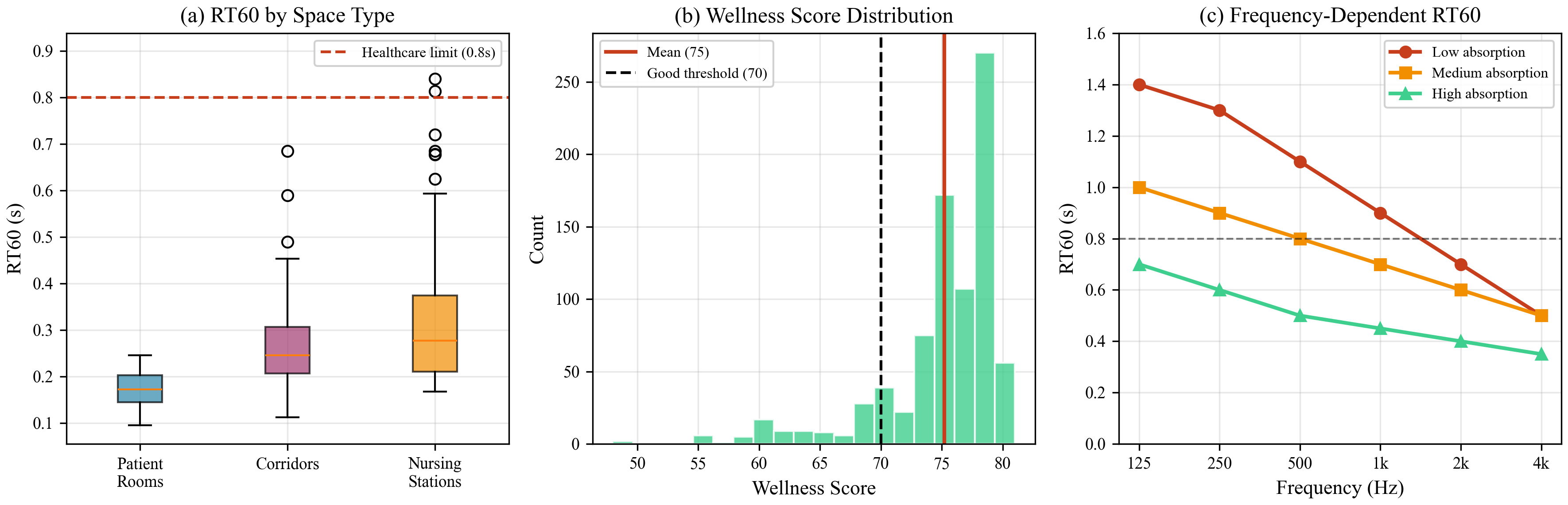}
    \caption{(a) Comparison of acoustic parameters across healthcare space types. (b) Wellness score distribution. (c) Octave-band RT60 patterns showing frequency-dependent absorption effects.}
    \label{fig:healthcare_study}
\end{figure*}

\subsection{Case Study: Healthcare Facility Design}

Hospital acoustics affect patient recovery, staff communication, and medical error rates. We evaluated impulse responses representing patient rooms, corridors, and nursing stations (Figure 5).
\begin{table}[htbp]
\centering
\caption{Summary of classroom acoustics case study results including sample sizes, compliance rates, and statistical comparisons.}
\label{tab:classroom-results}
\begin{tabular}{lc}
\toprule
\textbf{Metric} & \textbf{Value} \\
\midrule
Number of rooms & 335 \\
Room volume (m$^3$) & 241.3 $\pm$ 71.3 \\
RT60 (s) & 0.43 $\pm$ 0.20 \\
STI proxy & 0.79 $\pm$ 0.07 \\
ANSI compliance rate & 84.2\% \\
RT60-STI correlation & $r = -0.992$ \\
\bottomrule
\end{tabular}
\end{table}

\subsection{Case Study: Recording Studio Evaluation}

Professional recording environments demand exceptionally low reverberation and controlled frequency response. We analyzed RIRs with characteristics matching studio design targets (Figure 6).

\begin{figure*}[t]
    \centering
    \includegraphics[width=2.1\columnwidth]{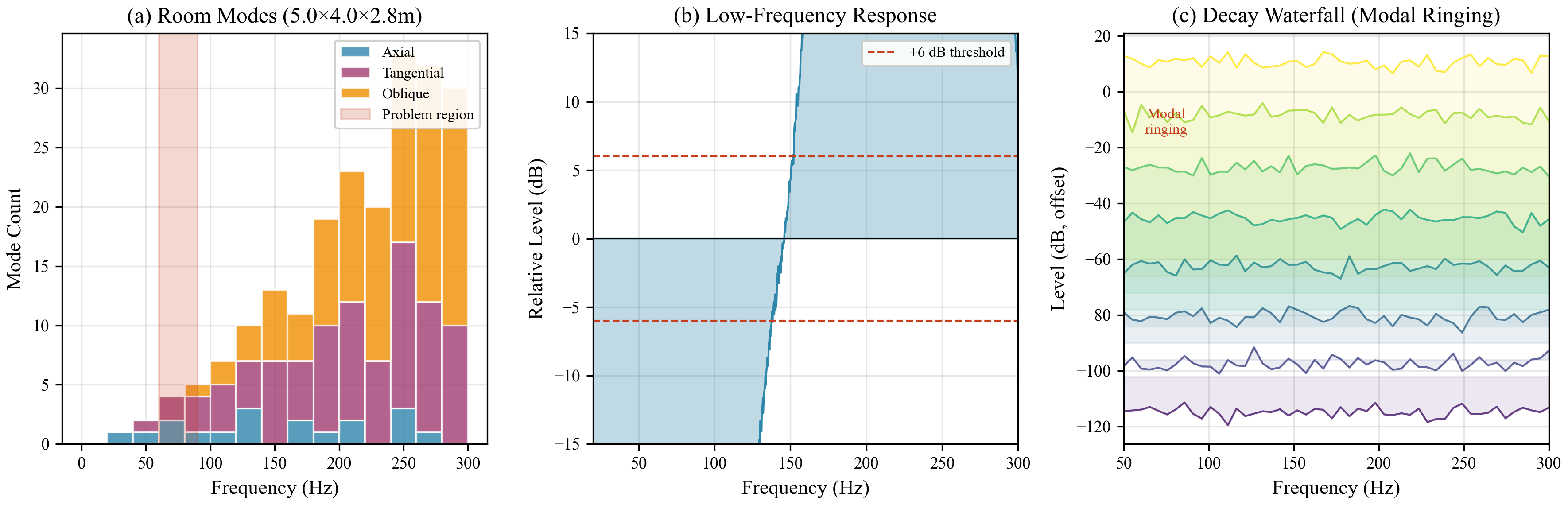}
    \caption{(a) Room mode distribution for a small studio room showing problematic clustering. (b) Frequency response showing bass buildup. (c) Waterfall plot revealing modal ringing at low frequencies.}
    \label{fig:studio_study}
\end{figure*}

The room mode analysis proved particularly valuable for identifying potential acoustic problems before physical construction.

\subsection{Performance Benchmarks}

We measured processing time for the complete analysis pipeline across impulse responses of varying lengths.
\begin{table}[t]
\centering
\small
\caption{Processing time (seconds) by RIR duration on consumer hardware (Intel i7, 16GB RAM).}
\label{tab:performance}
\begin{tabular}{lccc}
\toprule
\textbf{Component} & \textbf{1s RIR} & \textbf{5s RIR} & \textbf{10s RIR} \\
\midrule
Loading/preprocessing     & 0.000 & 0.001 & 0.003 \\
Broadband metrics          & 0.004 & 0.010 & 0.019 \\
Octave-band analysis       & 0.014 & 0.046 & 0.088 \\
Visualization generation   & 1.837 & 0.012 & 0.011 \\
\midrule
\textbf{Total}             & \textbf{1.855} & \textbf{0.069} & \textbf{0.121} \\
\bottomrule
\end{tabular}
\end{table}

\section{Discussion}
\label{sec:discussion}

\subsection{Limitations}

Several limitations should be acknowledged:

\textbf{STI Proxy}: The implemented STI calculation is a simplified proxy rather than the full IEC 60268-16 procedure. For formal compliance testing, users should employ certified measurement systems.

\textbf{Simulated Data}: The \dataset{} collection contains simulated rather than measured impulse responses. While geometric acoustic simulation captures many important phenomena, it may not fully represent complex real-world acoustic conditions including diffraction, air absorption at high frequencies, and non-uniform surface properties.

\textbf{Rectangular Room Assumption}: The image-source visualization and room mode calculations assume rectangular geometry. Non-rectangular spaces require more sophisticated treatment.

\textbf{Single-Point Analysis}: The current implementation analyzes single source-receiver pairs. Spatial variation of acoustic parameters within a room requires multiple measurement positions.

\subsection{Future Directions}

Several extensions are planned for future development:

\begin{itemize}
    \item Full IEC 60268-16 STI implementation with octave-band modulation transfer functions
    \item Support for measured binaural impulse responses with HRTF-based auralization
    \item Batch processing mode for analyzing multiple RIRs with comparative visualization
    \item Integration with acoustic simulation tools for predictive modeling
    \item Mobile-optimized interface for field measurements
\end{itemize}

\section{Conclusion}
\label{sec:conclusion}

This paper presented \system{}, an open-source web platform for comprehensive room impulse response analysis. The system implements twelve analysis modes covering temporal, spectral, and spatial acoustic characteristics with mathematical rigor grounded in international standards. The accompanying \dataset{} dataset provides thousands of simulated impulse responses with rich metadata for research and educational use.

By combining professional-grade acoustic analysis with an accessible web interface, \system{} aims to democratize room acoustics expertise. Architects can evaluate design alternatives without specialized training. Researchers can rapidly analyze large RIR collections. Educators can provide students with hands-on experience in acoustic measurement interpretation.

The platform is freely available at \url{https://huggingface.co/spaces/mandipgoswami/acoustivision-pro} with source code released under an open-source license. The \dataset{} and RIRMega Speech datasets are hosted on Hugging Face at the URLs provided in the introduction.

\section*{Acknowledgments}
We thank colleagues who reviewed the scripts and suggested diagnostic plots including the duration-RT60 scatter and features development.

\bibliographystyle{IEEEtran}

\begin{thebibliography}{99}

\bibitem{iso3382_1}
ISO 3382-1:2009, ``Acoustics -- Measurement of room acoustic parameters -- Part 1: Performance spaces,'' International Organization for Standardization, 2009.

\bibitem{iso3382_2}
ISO 3382-2:2008, ``Acoustics -- Measurement of room acoustic parameters -- Part 2: Reverberation time in ordinary rooms,'' International Organization for Standardization, 2008.

\bibitem{iso3382_3}
ISO 3382-3:2012, ``Acoustics -- Measurement of room acoustic parameters -- Part 3: Open plan offices,'' International Organization for Standardization, 2012.

\bibitem{ansi_s12_60}
ANSI/ASA S12.60-2010, ``Acoustical Performance Criteria, Design Requirements, and Guidelines for Schools,'' American National Standards Institute, 2010.

\bibitem{iec_60268_16}
IEC 60268-16:2020, ``Sound system equipment -- Part 16: Objective rating of speech intelligibility by speech transmission index,'' International Electrotechnical Commission, 2020.

\bibitem{schroeder1965}
M.~R. Schroeder, ``New method of measuring reverberation time,'' \textit{Journal of the Acoustical Society of America}, vol.~37, no.~6, pp.~1187--1188, 1965.

\bibitem{bradley1995}
J.~S. Bradley, ``Predictors of speech intelligibility in rooms,'' \textit{Journal of the Acoustical Society of America}, vol.~80, no.~3, pp.~837--845, 1986.

\bibitem{bradley1986}
J.~S. Bradley, ``Speech intelligibility studies in classrooms,'' \textit{Journal of the Acoustical Society of America}, vol.~80, no.~3, pp.~846--854, 1986.

\bibitem{beranek2004}
L.~L. Beranek, \textit{Concert Halls and Opera Houses: Music, Acoustics, and Architecture}, 2nd ed. New York: Springer, 2004.

\bibitem{houtgast1985}
T.~Houtgast and H.~J.~M. Steeneken, ``A review of the MTF concept in room acoustics and its use for estimating speech intelligibility in auditoria,'' \textit{Journal of the Acoustical Society of America}, vol.~77, no.~3, pp.~1069--1077, 1985.

\bibitem{schroeder1996}
M.~R. Schroeder, ``The `Schroeder frequency' revisited,'' \textit{Journal of the Acoustical Society of America}, vol.~99, no.~5, pp.~3240--3241, 1996.

\bibitem{odeon}
ODEON Room Acoustics Software, \url{https://odeon.dk/}.

\bibitem{catt}
CATT-Acoustic, \url{https://www.catt.se/}.

\bibitem{easera}
EASERA, \url{https://easera.afmg.eu/}.

\bibitem{python_acoustics}
Python Acoustics, \url{https://github.com/python-acoustics/python-acoustics}.

\bibitem{pyroomacoustics}
R.~Scheibler, E.~Bezzam, and I.~Dokmani\'{c}, ``Pyroomacoustics: A Python package for audio room simulation and array processing algorithms,'' in \textit{Proc. IEEE ICASSP}, 2018, pp.~351--355.

\bibitem{rew}
Room EQ Wizard, \url{https://www.roomeqwizard.com/}.

\bibitem{ace_challenge}
J.~Eaton, N.~D. Gaubitch, A.~H. Moore, and P.~A. Naylor, ``Estimation of room acoustic parameters: The ACE challenge,'' \textit{IEEE/ACM Trans. Audio, Speech, Language Process.}, vol.~24, no.~10, pp.~1681--1693, 2016.

\bibitem{mit_ir}
MIT Impulse Response Survey, \url{https://mcdermottlab.mit.edu/Reverb/IR_Survey.html}.

\bibitem{geometric_acoustics}
U.~P. Svensson, R.~I. Fred, and J.~Vanderkooy, ``An analytic secondary source model of edge diffraction impulse responses,'' \textit{Journal of the Acoustical Society of America}, vol.~106, no.~5, pp.~2331--2344, 1999.

\bibitem{wave_based}
S.~Bilbao, \textit{Numerical Sound Synthesis: Finite Difference Schemes and Simulation in Musical Acoustics}. Chichester, UK: Wiley, 2009.

\bibitem{vorlaender2007}
M.~Vorl\"{a}nder, \textit{Auralization: Fundamentals of Acoustics, Modelling, Simulation, Algorithms and Acoustic Virtual Reality}. Berlin: Springer, 2007.

\bibitem{gradio}
A.~Abid, A.~Abdalla, A.~Abid, D.~Khan, A.~Alfozan, and J.~Zou, ``Gradio: Hassle-free sharing and testing of ML models in the wild,'' in \textit{Proc. ICML Workshop on Human in the Loop Learning}, 2019.

\end{thebibliography}

\appendix

\section{Standards Compliance Thresholds}
\label{app:standards}

Table~\ref{tab:standards_full} lists the complete set of standards and thresholds implemented in \system{}.

\begin{table}[h]
    \centering
    \caption{Standards compliance thresholds implemented in \system{}.}
    \label{tab:standards_full}
    \begin{tabular}{lcc}
        \toprule
        \textbf{Standard/Space Type} & \textbf{RT60} & \textbf{STI} \\
        \midrule
        Classroom (ANSI S12.60) & $\leq$ 0.6 s & $\geq$ 0.60 \\
        Open Office (ISO 3382-3) & $\leq$ 0.8 s & $\geq$ 0.50 \\
        Private Office & $\leq$ 0.6 s & $\geq$ 0.55 \\
        Hospital Ward & $\leq$ 0.8 s & $\geq$ 0.60 \\
        Concert Hall & 1.5--2.5 s & -- \\
        Lecture Hall & $\leq$ 1.0 s & $\geq$ 0.55 \\
        Recording Studio & $\leq$ 0.4 s & -- \\
        Worship Space & 1.2--3.0 s & -- \\
        Restaurant & $\leq$ 0.9 s & -- \\
        Conference Room & $\leq$ 0.7 s & $\geq$ 0.55 \\
        \bottomrule
    \end{tabular}
\end{table}

\section{Octave-Band Filter Specifications}
\label{app:filters}

The octave-band analysis employs fourth-order Butterworth bandpass filters with the following characteristics:

\begin{table}[h]
    \centering
    \caption{Octave-band filter specifications.}
    \label{tab:filters}
    \begin{tabular}{ccc}
        \toprule
        \textbf{Center (Hz)} & \textbf{Lower (Hz)} & \textbf{Upper (Hz)} \\
        \midrule
        125 & 88 & 177 \\
        250 & 177 & 354 \\
        500 & 354 & 707 \\
        1000 & 707 & 1414 \\
        2000 & 1414 & 2828 \\
        4000 & 2828 & 5657 \\
        \bottomrule
    \end{tabular}
\end{table}

The cutoff frequencies follow the standard relationship $f_{\text{lower}} = f_c / \sqrt{2}$ and $f_{\text{upper}} = f_c \cdot \sqrt{2}$.

\section{Room Volume Effects}
\label{app:volume}
See Figure 7.

\begin{figure}[htbp]
    \centering
    \includegraphics[width=1.1\columnwidth]{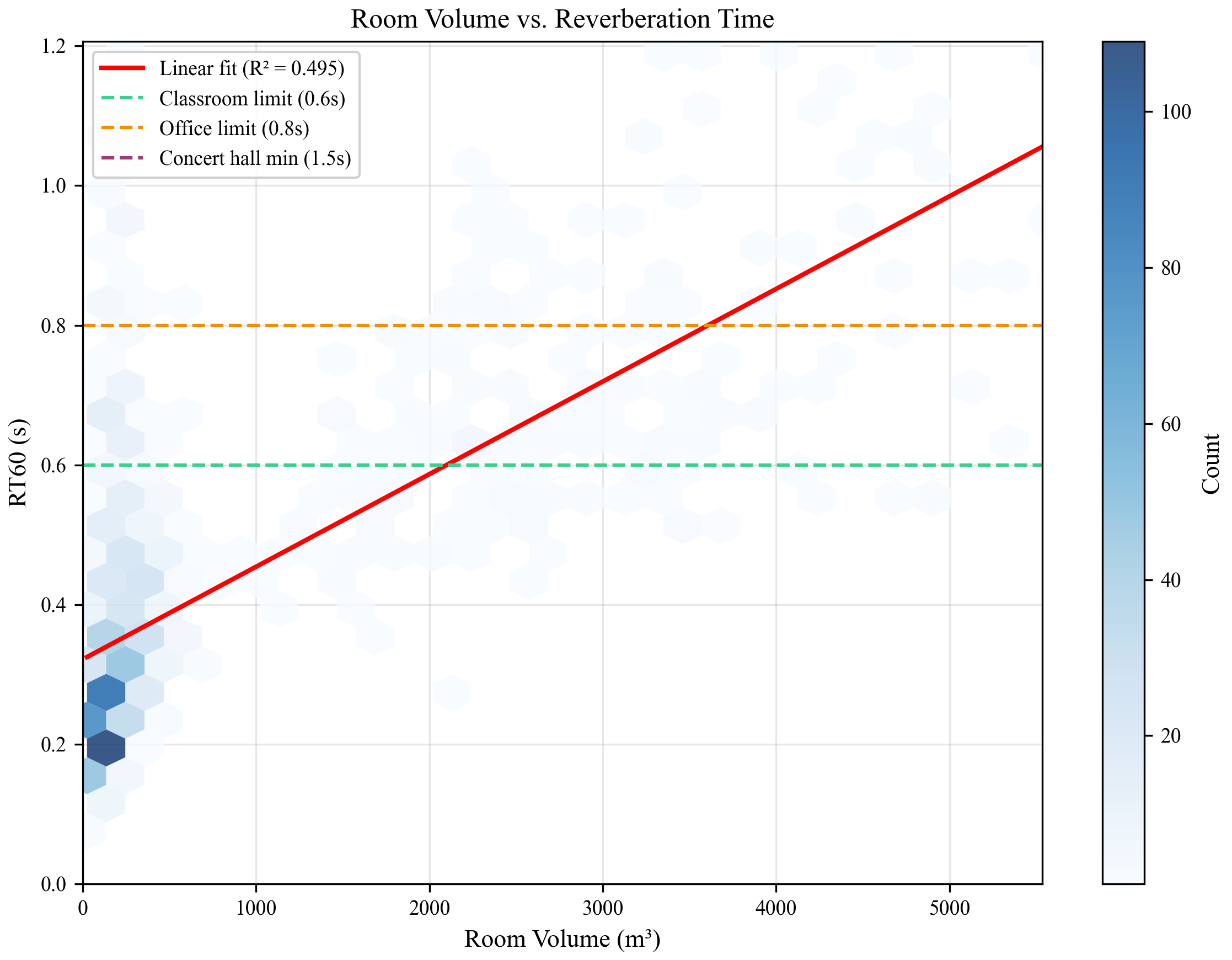}
    \caption{Room Volume Effects on RT60}
    \label{fig:volume}
\end{figure}

\section{Absorption Effects}
\label{app:absorption}
See Figure 8.

\begin{figure*}[t]
    \centering
    \includegraphics[width=2.0\columnwidth]{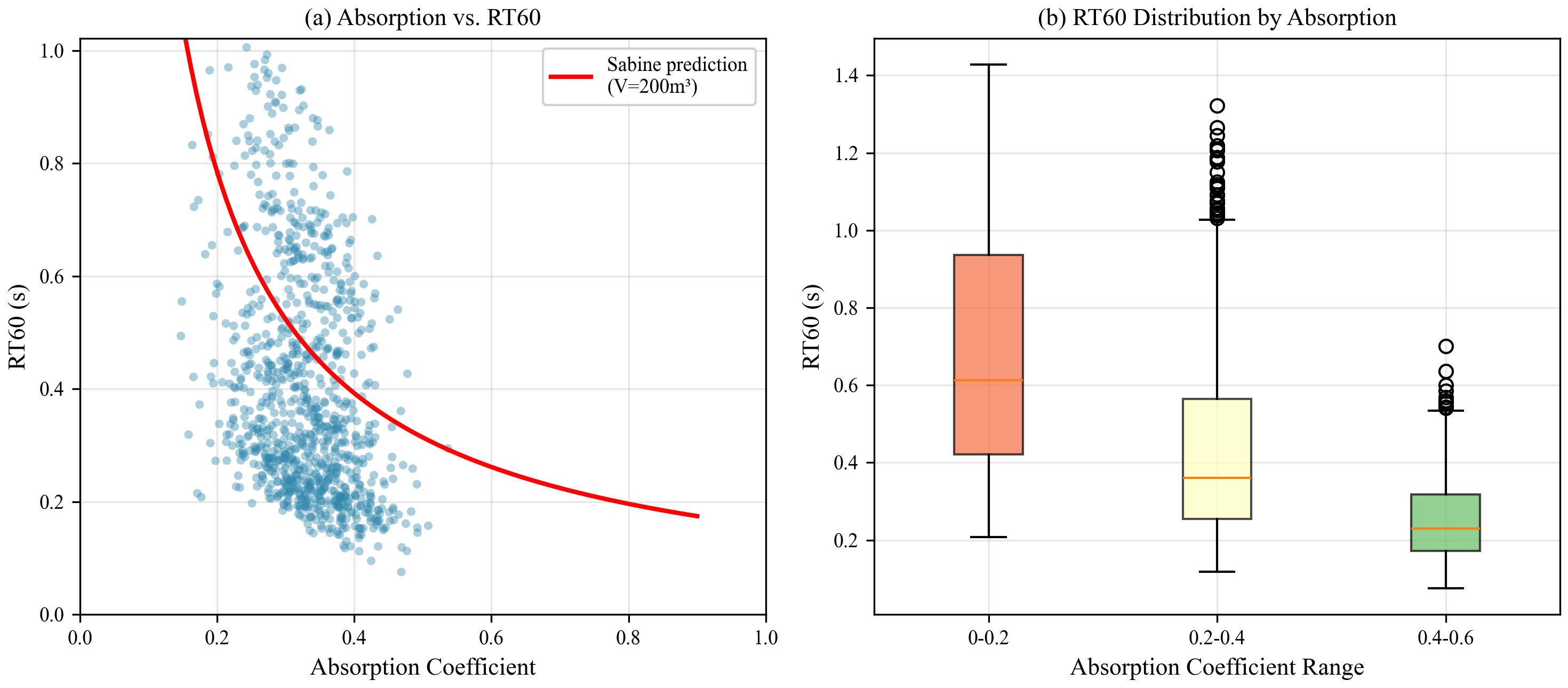}
    \caption{Effect of Absorption on RT60}
    \label{fig:absorption}
\end{figure*}

\section{Clarity Distribution}
\label{app:clarity}
See Figure 9.

\begin{figure*}[htbp]
    \centering
    \includegraphics[width=2.1\columnwidth]{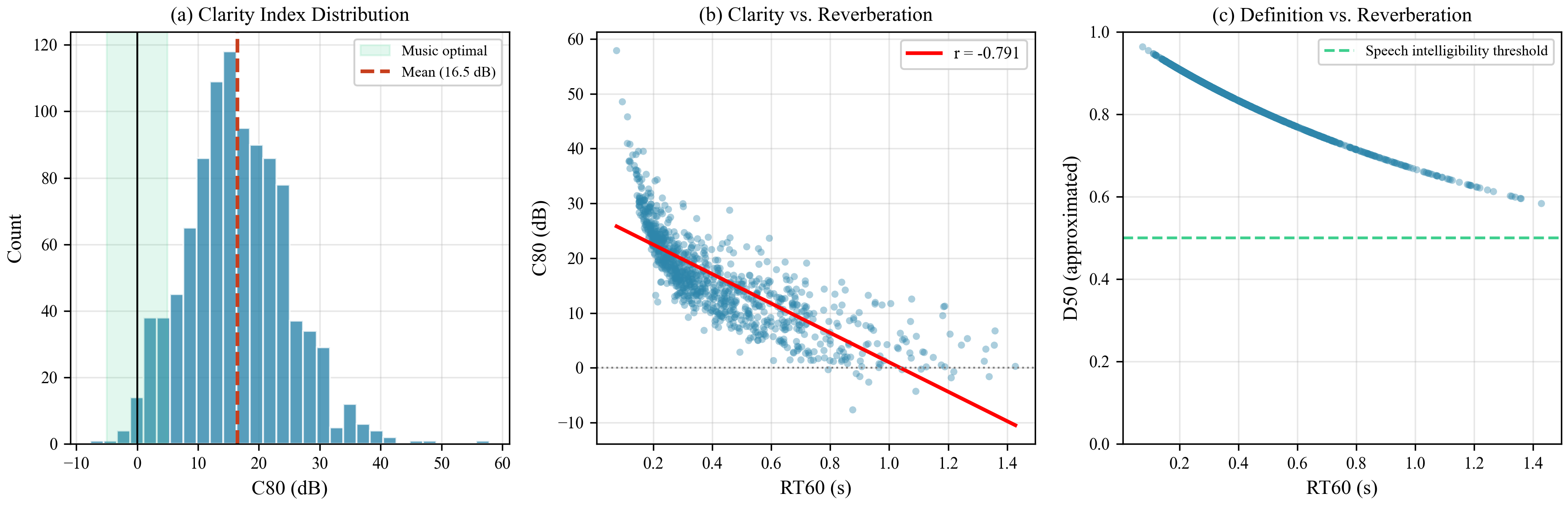}
    \caption{Clarity distribution and comparison to RT60}
    \label{fig:clarity}
\end{figure*}

\end{document}